\documentclass[%
 reprint,
superscriptaddress,
 amsmath,amssymb,
 aps,
 prl,
]{revtex4-1}

\usepackage{graphicx}
\usepackage{dcolumn}
\usepackage{bm}
\usepackage{color}
\usepackage{amsmath}

\begin{document}


\title{A Model for Excess Johnson Noise in Superconducting Transition-edge Sensors}



\author{Abigail Wessels}
\affiliation{University of Colorado, Boulder, CO 80309, USA.}
\affiliation{National Institute of Standards and Technology, Boulder, CO, 80305 USA}

\author{Kelsey Morgan}
\author{Daniel T. Becker}
\affiliation{University of Colorado, Boulder, CO 80309, USA.}
\affiliation{National Institute of Standards and Technology, Boulder, CO, 80305 USA}

\author{Johnathon D. Gard}
\affiliation{University of Colorado, Boulder, CO 80309, USA.}
\affiliation{National Institute of Standards and Technology, Boulder, CO, 80305 USA}

\author{Gene C. Hilton}
\affiliation{National Institute of Standards and Technology, Boulder, CO, 80305 USA}

\author{John A. B. Mates}
\affiliation{University of Colorado, Boulder, CO 80309, USA.}
\affiliation{National Institute of Standards and Technology, Boulder, CO, 80305 USA}

\author{Carl D. Reintsema}
\author{Daniel R. Schmidt}
\author{Daniel S. Swetz}
\affiliation{National Institute of Standards and Technology, Boulder, CO, 80305 USA}

\author{Joel N. Ullom}
\affiliation{University of Colorado, Boulder, CO 80309, USA.}
\affiliation{National Institute of Standards and Technology, Boulder, CO, 80305 USA}

\author{Leila R. Vale}
\affiliation{National Institute of Standards and Technology, Boulder, CO, 80305 USA}

\author{Douglas A. Bennett}
\email[]{Douglas.Bennett@nist.gov}
\thanks{Contribution of a U.S. government agency, not subject to copyright.}
\affiliation{National Institute of Standards and Technology, Boulder, CO, 80305 USA}



\date{\today}

\begin{abstract}

Transition-Edge Sensors (TESs) are two-dimensional superconducting films used to detect energy or power.  TESs are voltage biased in the resistive transition where the film resistance is both finite and a strong function of temperature.  Electrical noise is observed in TESs that exceeds the predictions of existing noise theories. In this manuscript, we describe a model for the unexplained excess noise based on the dynamic resistance of the TES and noise mixed down from frequencies around the Josephson oscillations. We derive an expression for the power spectral density of this noise and show that its predictions match measured data.  

\end{abstract}


\maketitle



%



Transition-edge sensors (TESs) are the core sensor technology for an increasing variety of instruments at cutting-edge scientific facilities \cite{barret2018athena, doriese2017practical, galitzki2018simons} and measurements in a variety of different applications including cosmology \cite{spilker2018fast, miller2018massive},  quantum information \cite{giustina2015significant} \cite{shen2018randomness}, neutrino mass measurements \cite{alpert2015holmes}, exotic atom experiments \cite{okada2014high}, and x-ray metrology \cite{fowler2017reassessment}. A TES is a two-dimensional superconducting film that is voltage-biased in the phase transition between the normal and superconducting states.  A TES can be used to very precisely measure small changes in temperature due to the absorption of single photons and particles or the average power deposited by many photons. TESs can be optimized to achieve excellent energy resolution or noise equivalent power. As demanding applications such as x-ray spectroscopy at the LCLS-II \cite{li2018tes} push transition-edge sensors (TESs) to even better energy resolution, it is critical to understand all their potential noise sources. Noise that is commonly observed in TESs in excess of the noise predicted by known mechanisms limits the optimization of TESs for challenging applications. In this letter, we propose a mechanism for the unexplained excess noise, derive an expression for its amplitude, and compare the predictions of the expression to measured noise values.

As soon as TES were carefully compared to noise theories for bolometers\cite{mather1982bolometer} and microcalorimeters \cite{moseley1984thermal}, electrical noise in excess of these models was consistently observed \cite{ullom2004suppression, kinnunen2008reducing, smith2013implications}. For a more detailed discussion on the observation of excess noise see reviews by Irwin and Hilton \cite{irwin2005transition} and Galeazzi\cite{galeazzi2010fundamental}. Noise sources in a TES can be divided into (1) noise sources internal to the TES, (2) noise sources from the circuit in which the TES is embedded such as Johnson noise in the bias resistor and noise in the readout amplifier, and (3) noise from the external environment such as RF-pickup, stray photon arrivals, and fluctuations in the temperature bath. The second and third categories of noise from this list are well understood and are not the source of the excess noise discussed here. 

\begin{figure}
\centering
\includegraphics[width=3.4in]{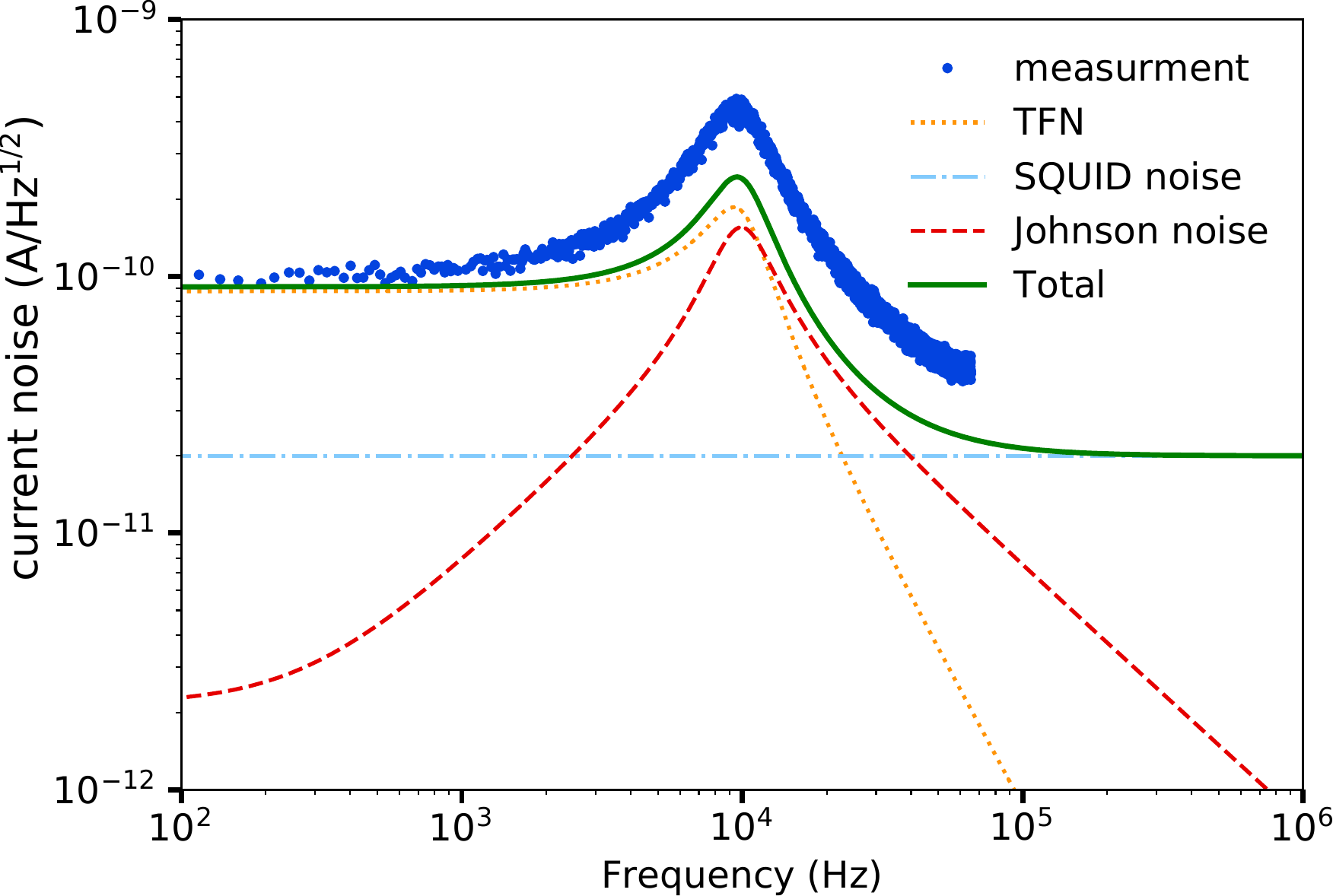}
\caption{Current noise $(A/\sqrt{Hz})$ as a function of frequency in TES. Blue points are the measured data. The lines are predictions for the thermal fluctuation noise (orange dotted line), SQUID noise (cyan dashed-dotted line), Johnson noise including the $(1+2\beta)$ term of \cite{irwin2006thermodynamics} (red dashed line), and the quadrature sum of the noise components (green solid line). }
\label{fig:m_noise}
\end{figure}

The main internal noise sources that have been observed in TESs are the thermal fluctuation noise (TFN) between the TES and the heat bath, and the voltage fluctuations due to Johnson noise from the resistance of the TES \cite{irwin2005transition}. The power fluctuations due to thermal fluctuation noise are expressed as $S_{P_{TFN}} = 4 k_B T^2 G  F_L$ where the $T$ is the temperature of the TES, $G$ is the thermal conductance to the bath, and $F_L$ is a unitless function that depends on the thermal conductance exponent and on whether phonon transport to the TES is specular or diffuse. $S_{P_{TFN}}$ is frequency independent. The current in a TES is usually measured by a SQUID ammeter, and it is convenient to convert all types of noise into current noise as measured by the SQUID. To convert the thermal power fluctuations to current fluctuations in the TES, we multiply by the the power-to-current responsivity of the TES. Since the power-to-current responsivity is frequency dependent, the resulting current fluctuations are also frequency dependent. The voltage fluctuations due the Johnson noise of the TES are expressed as  $S_V = 4 k_B T R$ where $R$ is the resistance of the TES at its operating point. $S_V$ is independent of frequency. To convert the voltage noise to a current noise as seen in the measured current, we multiply by the internal admittance of the TES circuit. 

Figure \ref{fig:m_noise} shows predicted current noise components due to TFN and the TES Johnson noise while the points are measured data for a typical TES. The measured data shows noise in excess of the noise predictions with the same frequency dependence as the TES Johnson noise. This excess noise, often referred to as ``unexplained noise'', is the noise that is not understood and the focus of this letter. It is usually parameterized as an additional factor of $(1+M^2)$ multiplying the expected Johnson noise\cite{ullom2004suppression}.

This noise should not be confused with an additional noise source sometimes observed due to internal thermal fluctuations between distributed heat capacities inside the TES \cite{hoevers2000thermal, gildemeister2001model}. The frequency dependence of internal thermal fluctuation noise (ITFN) can sometimes mimic that of the Johnson noise. However, IFTN is well understood, such that when the TES parameters are carefully measured, it can usually be definitively separated from the unexplained excess noise \cite{wakeham2019thermal}. Also IFTN can usually be suppressed below all other noise sources by use of low resistance TESs. 

In 2006, Irwin  \cite{irwin2006thermodynamics} predicted an enhancement of the Johnson noise based on an analysis of a simple nonlinear resistive calorimeter operated out of equilibrium. This analysis predicted that the noise level in a TES is equal to the Johnson noise due to the TES resistance at its operating point multiplied by a factor of $(1+2\beta_I)$ where $\beta_I$ is the logarithmic current sensitivity of the TES defined as
\begin{equation}
\beta_I =  \left. \frac{\partial \log R}{\partial \log I} \right|_{T} = \frac{I}{R} \left. \frac{\partial R}{\partial I} \right|_{T} . 
\end{equation} 
The parameters $R$, $\beta_I$, and $\alpha_I$, the logarithmic temperature sensitivity defined as 
\begin{equation}
\alpha_I =  \left. \frac{\partial \log R}{\partial \log T} \right|_{I} = \frac{T}{R} \left. \frac{\partial R}{\partial t} \right|_{I},
\end{equation} 
are commonly used to describe the $R(I,T)$ surface in the small signal limit\cite{irwin2005transition}.  

In some scenarios the  $(1+2\beta_I)$ term has reasonably predicted measured noise. In many other scenarios, especially low in the transition and for devices with high $\alpha_I$and $\beta_I$, the $(1+2\beta_I)$ expression dramatically underpredicts the observed noise \cite{jethava2009dependence}. The factor of $(1+2\beta)$ is included in the predicted Johnson noise in Fig.\ \ref{fig:m_noise}.  It is now usual to refer to the excess noise as the noise above the predictions of the $(1 + 2 \beta_I)$ term.  The $M$ parameter is then defined by
 \begin{equation}
 S_V (0) = 4 k_B T R (1 + 2 \beta_I) (1+M^2).
 \end{equation} 
In Fig.\ \ref{fig:m_noise}, an additional $M=2.7$ is required to make the predicted noise consistent with the measured data.

A number of explanations for the excess Johnson noise have been proposed including fluctuations due to vortex dynamics \cite{fraser2004nature}, fluctuations in the superconducting phase boundary \cite{luukanen2003fluctuation}, fluctuations in the superconducting order parameter \cite{seidel2004intrinsic}, and by percolation models \cite{bagliani2009study}. However none of these mechanisms give quantitative predictions consistent with the measured dependencies of the excess electrical noise. 
Despite more than a decade of theoretical and experimental effort, the magnitude of the unexplained noise in TESs is still not understood. 

One obstacle to understanding the excess noise is that there is no consensus model to predict the resistance of TESs biased in the superconducting transition. As implied by a non-zero $\beta_I$, the resistance in a TES is not only a function of temperature, but also a function of current in the film. Any model that attempts to describe the resistance of the TES must include a mechanism that describes the current dependence. Two such models have been proposed, one based on the observation of Josephson junction-like weak-link behavior in TESs \cite{sadleir2010longitudinal} and the other based on intermediate resistance states formed by phase-slips lines  \cite{irwin1998thermal, bennett2014phase}. Based on observations of weak link behavior, Kozorezov et al. \cite{kozorezov2011modelling} proposed modeling TESs using the resistively-shunted junction (RSJ) model. Despite reasonably describing some qualitative features of TESs, the model is not able to successfully predict measured values of $\alpha_I$ and $\beta_I$. An empirical model based on the phase-slip line mechanism, known as the two-fluid model  \cite{irwin1998thermal, bennett2012two}, has been used to predict $\alpha_I$ and $\beta_I$ over some fraction of the transition \cite{morgan2017dependence, pappas2014optical}, but relies on two empirical parameters. 

Interestingly, both of these proposed resistance mechanisms predict an induced oscillating supercurrent at the Josephson frequency ($\omega_J = 2 e V / \hbar$). Additionally, both mechanisms imply a non-linear resistance for the TES at least over some range of parameter space. Together, the high frequency oscillations and the non-linear nature of the resistance are suggestive that noise could be mixed down from higher frequencies as is commonly observed in dc-SQUIDs and was studied extensively in weak-link Josephson junctions.

 The amplitude the of Johnson noise mixed down to low frequencies in weak-link junctions was first calculated by Likharev and Semenov  \cite{likharev1972fluctuation} within the context of the RSJ model. In this formulation the only intrinsic source of fluctuations is the normal current. The supercurrent is not a source of fluctuations since the superconducting condensate is an ordered set of electron pairs, which cannot fluctuate independently. Therefore, the normal current passing through the resistive shunt is the source of all the Johnson noise within this model. In the theory of Likharev and Semenov, the spectral density of the voltage noise at frequency $\omega$ is determined by both the voltage noise at that frequency and the noise mixed-down to $\omega$ by noise near the Josephson frequency ($\omega_J$) and its harmonics as
 \begin{equation}
 S_V (\omega) =  \sum_k \left|  Z_k \right|^2  S_I \left( k \omega_J - \omega  \right) \label{psd_z}
 \end{equation}
 where $S_I$ is the spectral density of the current noise and $\left|  Z_k \right|$ are the Fourier coefficients of the junction impedance, defined as $V(\omega) = \sum_k Z_k I (\omega - k \omega_J )$. For frequencies small compared to $\omega_J$ only the $k=-1,0,1$ harmonics of the Josephson current are relevant \cite{likharev1972fluctuation} and the spectral density of the voltage fluctuations at zero frequency is given by
\begin{equation}
S_V (0) = R_d^2 S_I(0) + 2 \left| Z_1 \right|^2 S_I(\omega_J)
\end{equation}
 where $R_d$ is the dynamic resistance of the junction defined as $R_d = \partial V / \partial I$.
 
By calculating the Fourier coefficients of the junction impedance from differential equations of the RSJ model, Likharev and Semenov were able to derive the power spectral density of the voltage noise across the weak link at frequencies well below the Josephson frequency as 
 \begin{equation}
 S_V (0) =   4 k_B T \frac{R_d^2}{R_n} \left[ 1 + \frac{1}{2} \left( \frac{I_c}{I}  \right)^2  \right]. \label{psd_rd_rsj}
 \end{equation}
where $R_n$ is the shunt resistance across the junction, $I$ is the total current in the junction, and $I_c$ is the critical current of the junction. The second term is largest at $I=I_c$ and is reduced as the current is increased above $I_c$. Koch et al.\ \cite{koch1980quantum} later generalized this theory to include quantum effects due to noise from zero-point current fluctuations in the shunt resistor. However, at the voltage biases used in TESs, the quantum effects are obscured by thermal fluctuations and the quantum treatment reduces to Eq.\ \ref{psd_rd_rsj}. Under the assumption that a TES can be described by the RSJ model, we can substitute $R_d = R (1+ \beta_I)$ to rewrite the spectral density of the voltage fluctuations into parameters more usual for TESs as
 \begin{equation}
 S_V (0) =   4 k_B T \frac{R^2}{R_n} \left(1 + \beta_I \right)^2 \left[ 1 + \frac{1}{2} \left( \frac{I_c}{I}  \right)^2  \right]. \label{psd_beta_rsj}
\end{equation}

Inspired by the weak-link effects observed in TESs, Kozorezov et al.\ \cite{kozorezov2012electrical} applied the Smoluchowski equation approach of Coffey et al.\ \cite{coffey2008smoluchowski} in an attempt to explain the excess noise in TESs. The Smoluchowski equation approach is similar to the RSJ model except it takes into consideration thermal fluctuations.  However, in order to reach a calculable expression, Kozorezov et al.\ were forced to use an approximation and the resulting spectral density was the the same as Eq. \ref{psd_beta_rsj}. They then went on to use the RSJ form of the resistance $(R/R_n)^2 =1 - (I/I_c)^2 $ to write the power spectral density of the noise as 
\begin{equation}
 S_V (0) \approx 4 k_B T R (1 + \beta_I)^2 \left[ \frac{3}{2} \frac{R}{R_n} \left(1 - \frac{1}{3} \left( \frac{R}{R_n} \right)^2 \right) \right]. \label{psd_alex}
 \end{equation}
Later in the text we compare Eq.\ \ref{psd_alex} to measured data and show that it dramatically underpredicts the observed noise.

Since the transition shape in most TESs is not simply predicted by either the RSJ model or the two-fluid model, we would like a theory for the mixed-down noise that is independent of how closely the dynamic resistance of the device is described by one of the transition models. The first term of Eq.\ \ref{psd_rd_rsj} describes the power spectral density of the current fluctuations at low frequencies that is transformed into voltage fluctuations by multiplying by the local slope of the IV curve. The second term describes how Johnson noise at frequencies near the Josephson frequency interacts with the oscillations at the Josephson frequency to mix noise to low frequencies via the non-linearity of the resistance. 

What is required for TESs is a general expression for the second term of Eq.\ \ref{psd_rd_rsj} that does not rely on the specific form of the RSJ model. Kogan and Nagaev  \cite{kogan1988fluctuation} derived a general form for this expression. More recently the expression they derived was compared to experimental data for a weak-link junction by Lhotel et al.\  \cite{lhotel2007divergence}. The relevant Fourier coefficients of the junction impedance are
\begin{equation}
\left| Z_1 \right|^2 = - \frac{V}{4} \frac{\partial R_d}{\partial I}.
\end{equation}
giving the spectral density of the low frequency noise for arbitrary current-phase relationship as
\begin{equation}
S_V (0) = R_d^2 \left[ S_I(0) -  \frac{V}{2 R_d^2} \frac{\partial R_d}{\partial I} S_I(\omega_J) \right] \label{psd_general}. 
\end{equation}
In the context of a TES, $\partial R_d / \partial I$ must be evaluated at a constant temperature.

By substituting the dynamic resistance from the RSJ model and its derivative with respect to current into Eq. \ref{psd_general}, we recover Eq.\ \ref{psd_rd_rsj}. For the two-fluid model the dynamic resistance is a constant ($\partial R_d / \partial I=0$), and the second term of Eq.\ \ref{psd_general} is zero. Even though the phase-slip mechanism implies oscillations at the Josephson frequency, in the simplified form of two-fluid model, no noise is mixed down to low frequencies.

Using the definition of $\beta_I$ we can rewrite Eq.\ \ref{psd_general} to get
 \begin{equation}
S_V (0) = R_d^2 \left[ S_I(0) -  \frac{1}{2} \frac{\beta_I}{(1+\beta_I)^2} \frac{\partial R_d}{\partial R} S_I(\omega_J) \right]. \label{noise_noass}
\end{equation}
Equation \ref{noise_noass} makes no assumptions about the transition model except that there are Josephson oscillations being mixed with a device with some dynamic resistance. In order to compare Eq.\ \ref{noise_noass} to data we need to make an assumption about the form of $S_I$. If we take the RSJ model literally then we would assume that we have a normal resistance $R_n$ that shunts the junction and $S_I(0) = 4 k_B T / R_n$. In the TES there is no shunt; the resistance is intrinsic to the TES. When $S_I(0) = 4 k_B T / R_n$ is used, Eq.\ \ref{noise_noass} obviously under predicts the Johnson noise. As we will see in the remainder of this letter, assuming $S_I(0) = 4  k_B T / R$ does a significantly better job fitting the data. This same observation was made by Lhotel et al.\ \cite{lhotel2007divergence} in their comparison with long diffusive superconductor -- normal-metal -- superconductor junctions.

Taking $S_I(0) = 4  k_B T / R$, Eq.\ \ref{noise_noass} becomes
 \begin{equation}
S_V (0) = 4 k_B T R \left(1+\beta_I \right)^2 \left[ 1 -  \frac{1}{2} \frac{\beta_I}{(1+\beta_I)^2} \frac{\partial R_d}{\partial R} \right]. \label{noise_R}
\end{equation}
For comparison with data, we also write down the results of applying the RSJ model,
\begin{equation}
S_V (0) = 4 k_B T R \left(1+ \frac{5}{2} \beta_I + \frac{3}{2} \beta_I^2 \right), \label{noise_RSJ}
\end{equation}
and for the two-fluid model
\begin{equation}
S_V (0) = 4 k_B T R \left(1+ 2 \beta_I +  \beta_I^2 \right) \label{noise_TF}.
\end{equation}

\begin{figure}
\centering
\includegraphics[width=3.4in]{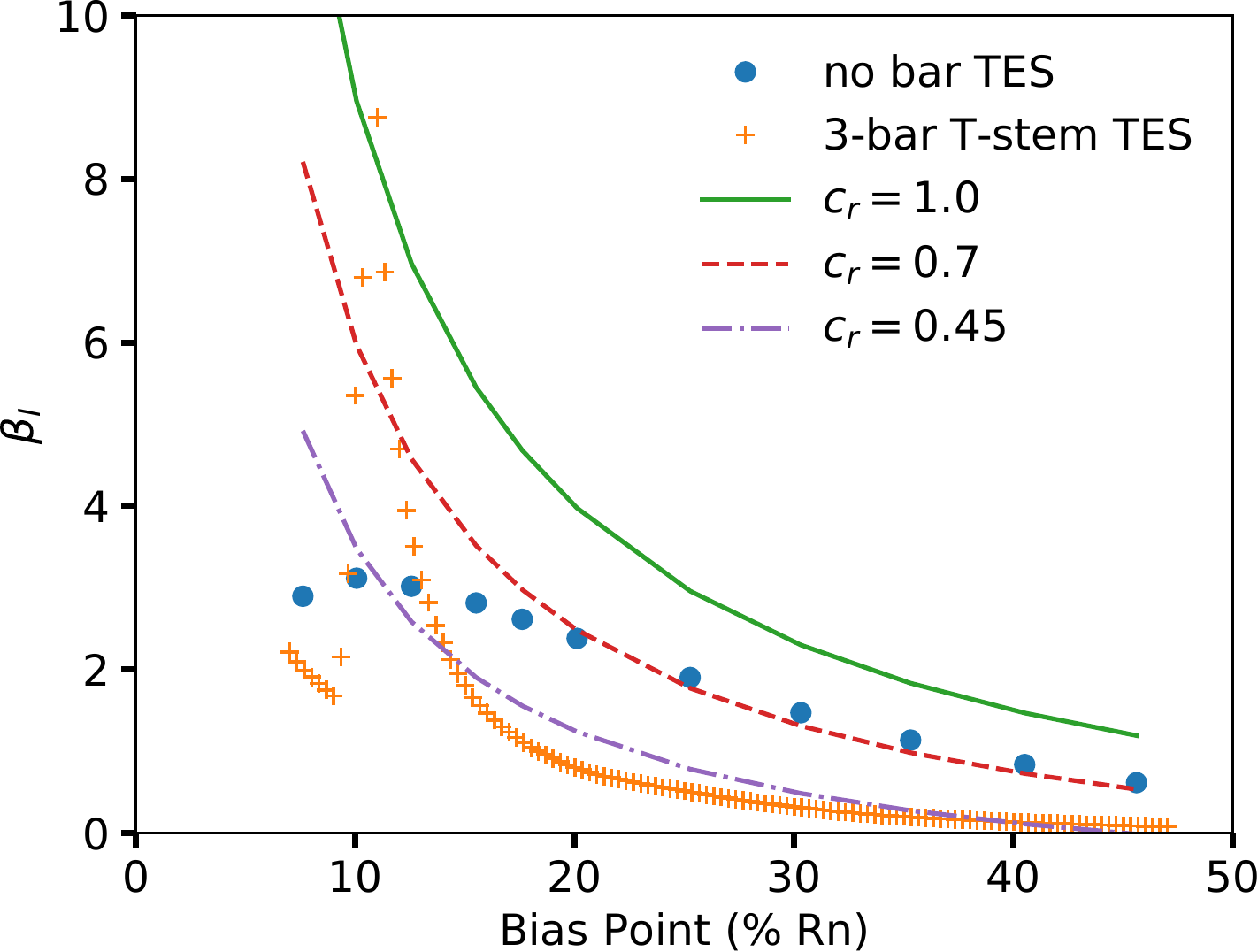}
\caption{Measured $\beta_I$ for no bar TES (blue points) and 3-bar T-stem TES (orange crosses) from Wakeham et al.\ \cite{wakeham2018effects}. The dashed lines are the two-fluid model predictions of $\beta_I$ for $c_R=1,0.7,0.45$.}
\label{fig:data_beta}
\end{figure}

We want to compare Eqs.\  \ref{noise_R}, \ref{noise_RSJ}, \ref{noise_TF} with data sets where $\beta_I$ and $M$ are known. The ideal data set would have  $\beta_I$ densely sampled around the $R$ values where we compare theory with measured noise.  Wakeham et al.\ \cite{wakeham2018effects} performed careful measurements of the $M$ parameter on well characterized 140 $\mu$m square TESs based on MoAu bilayers with gold banks parallel to the direction of current flow. One of the devices is a so-called T-stem with 3 normal metal stripes. The other TES has no stripes. The measurements of $\beta_I$ and $M$ as function of bias point were performed at 55 mK.

Figure \ref{fig:data_beta} shows the measured $\beta_I$ values for the no bar and 3-bar TESs along with lines corresponding to the predictions of $\beta_I$ versus bias point for the two-fluid model with $c_r = 1,0.7,0.45$. $c_r$ is one of the two empirical parameters in the two fluid model \cite{bennett2012two}. The dashed lines are useful in understanding how $R_d$ is changing with bias point expressed as \% $R_n$. For example, the no bar device follows the $c_r=0.7$ lines above 20\% $R_n$ implying a constant $R_d$ over this region. For a decrease in bias point, if the increase in $\beta_I$ is larger than the predictions of the two-fluid model, Eq.\ \ref{noise_R} predicts additional noise.

 \begin{figure}
\centering
\includegraphics[width=3.4in]{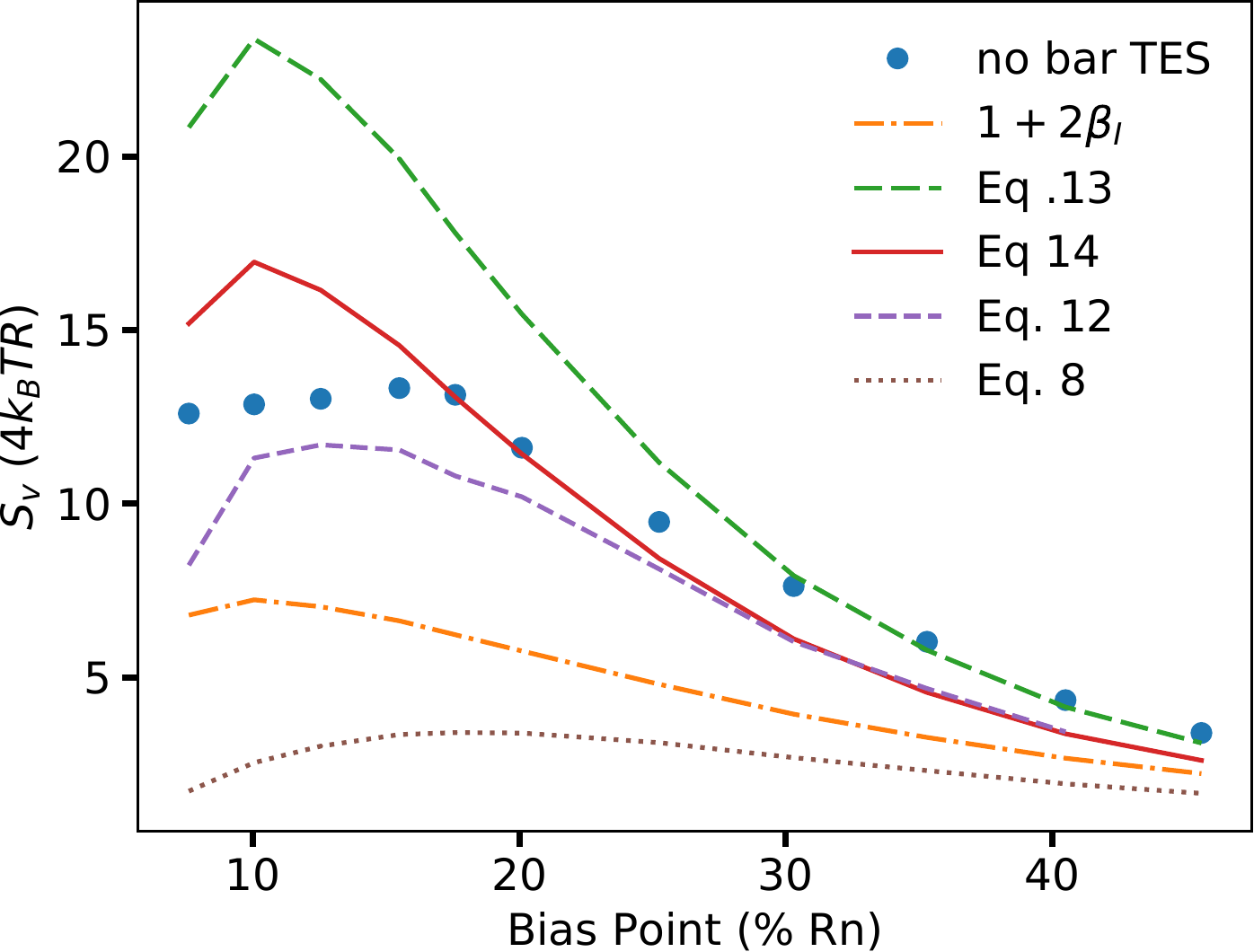}
\caption{The measured noise (blue points) as a function of bias point for the no bar TES. The orange dot-dashed line is the prediction of the usual $(1+2 \beta_I)$ term.  The other lines are the predicted Johnson noise using the two-fluid model (red) Eq.\  \ref{noise_TF},  RSJ model (green long dashed) Eq.\  \ref{noise_RSJ},  Eq.\ \ref{noise_R} (purple dashed) and Eq. \ref{psd_alex} (brown dotted).}
\label{fig:nobar_noise}
\end{figure}

 We use the measured values of $M^2$ and $\beta_I$ from \cite{wakeham2018effects} to calculate the total measured noise with frequency dependence consistent with Johnson nose. This gives the power spectral density of this component of the current noise in units of $4 k_B T R$. A value of one would be the usual Johnson noise spectral density for a resistance $R$ at temperature $T$.  For the no bar device, the extracted noise as a function of bias point in \% $R_n$ is shown as the blue circles in Fig.\ \ref{fig:nobar_noise}. The orange line shows what the prediction of the theory of \cite{irwin2006thermodynamics} where the Johnson noise is enhanced by $(1+2 \beta_I)$. It clearly under predicts the noise at all bias points. This trend is often observed in data with higher $\beta_I$. Eq.\ \ref{psd_alex}, shown as the brown dotted line, also under predicts all bias points. The green line is the prediction of Eq.\ \ref{noise_RSJ} and the red line is the prediction of  Eq.\ \ref{noise_TF}. Both do a better job of predicting the noise than the $1+2 \beta_I$ expression. Finally, the purple dashed line is the prediction from Eq.\ \ref{noise_R}. 

The $\partial R_d / \partial R $ term in Eq.\ \ref{noise_R} is estimated using the measured $\beta_I$ data from neighboring points in Fig. \ref{fig:data_beta} to calculate the rate of change of $\beta_I$. Since the neighboring points are at slightly different temperatures, this is only an estimate of $\partial R_d / \partial R $. This approximation gets worse for bias points that are further apart or in regions of especially high $\alpha_I$. The ideal data set would include $\beta_I$ measured at slightly different bath temperatures. Then the thermal model of the TES could be used to find which resistances are at the same temperature for the different bath temperatures and calculate the change in $\beta_I$ for a small change in $R$ while at fixed temperature.

The $R_d$ of the no bar TES is roughly flat above 25 \% $R_n$ as expected from  how well the two-fluid model matched the data over this region. Below 25 \% $R_n$, the $R_d$ is dropping. The resulting positive slope to $R_d$ causes the second term of Eq.\ \ref{noise_R} to be negative actually decreasing the noise compared to the prediction of a linear TES. We are currently exploring if an absolute value of this term was left out in the derivation of Kogan and Nagaev\cite{kogan1988fluctuation}.

\begin{figure}
\centering
\includegraphics[width=3.4in]{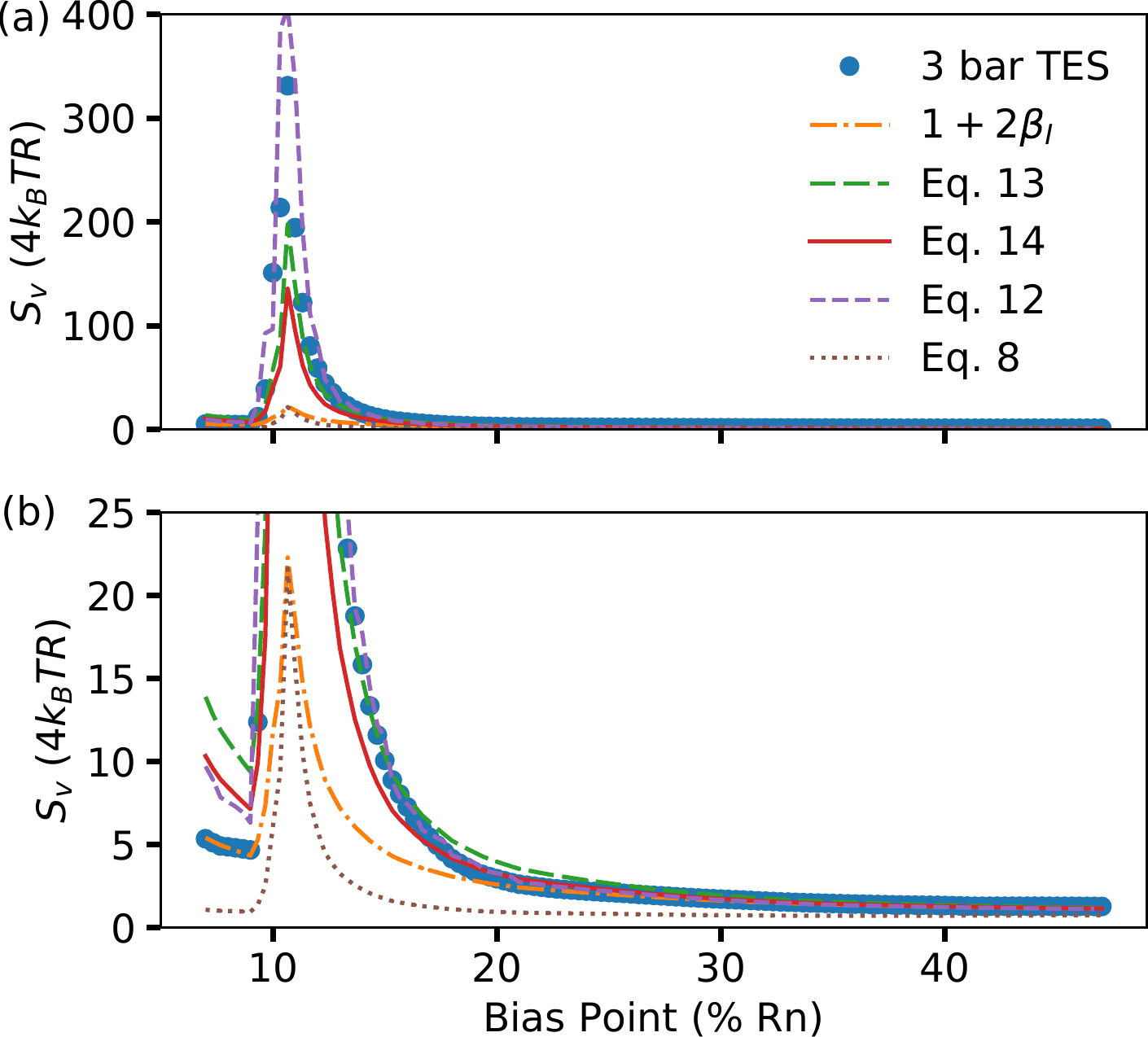}
\caption{ (a) The measured noise (blue points) as a function of bias point for the 3-bar T-stem TES. The orange dot-dashed line is the prediction of the usual $(1+2 \beta_I)$ term.  The other lines are the predicted Johnson noise using the two-fluid model (red) Eq.\  \ref{noise_TF},  RSJ model (green) Eq.\  \ref{noise_RSJ}, Eq.\ \ref{noise_R} (purple), and Eq. \ref{psd_alex} (brown dotted). (b) A zoomed version of (a) to compare noise away from the peak of the spiked noise feature.
}
\label{fig:3bar_noise}
\end{figure}

Figure \ref{fig:3bar_noise}a shows the measured noise (blue point) as a function of bias point for the 3-bar T-stem TES. Figure \ref{fig:3bar_noise}b is a zoomed in version of Fig.\ \ref{fig:3bar_noise}a to better show the data away from the peak of the noise spike. The orange line is the prediction of the usual $(1+2 \beta_I)$ term. The green line is the prediction of Eq.\ \ref{noise_RSJ} and the red line is the prediction of  Eq.\ \ref{noise_TF}. Both over predict the noise in some regions, i.e. around 22 \% $R_n$, but both do a better job near the spike in $\beta_I$. The purple dashed line is once again the prediction from Eq.\ \ref{noise_R}. Equation \ref{noise_R} appears to do an excellent job matching the measured noise. The only region it appears to not be a very good match is below 10 \% $R_n$

\begin{figure}
\centering
\includegraphics[width=3.4in]{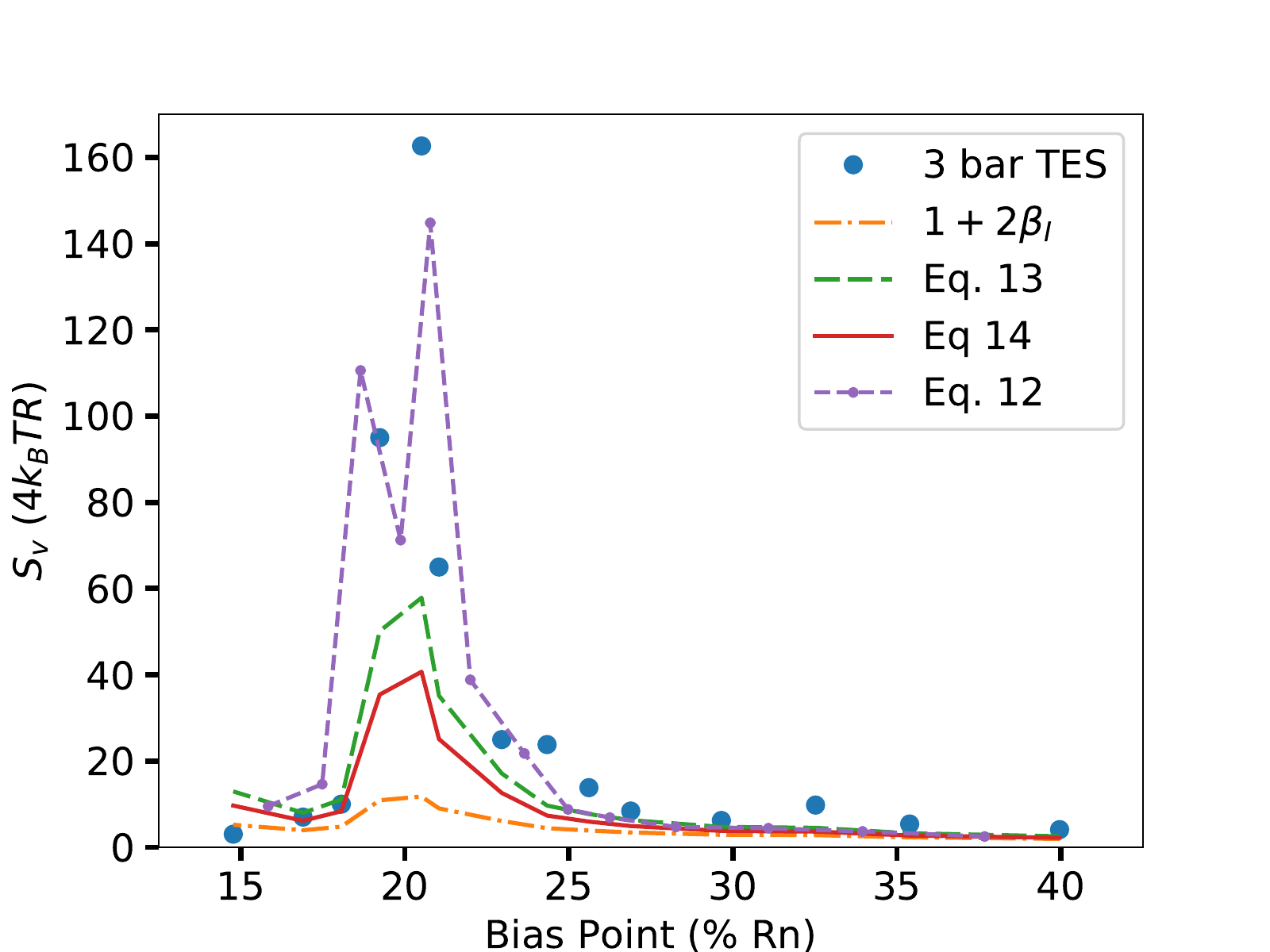}
\caption{The measured noise (blue points) as a function of bias point for a 124 $\mu$m 2-bar MoCu TES. The orange dot-dashed line is the prediction of the usual $(1+2 \beta_I)$ term.  The other lines are the predicted Johnson noise using the two-fluid model (red) Eq.\  \ref{noise_TF},  RSJ model (green) Eq.\  \ref{noise_RSJ}, and Eq.\ \ref{noise_R} (purple).
}
\label{fig:r11_noise}
\end{figure}

To compare Eqs.\ \ref{noise_R}-\ref{noise_TF} to a TES of a different bilayer material, we measured noise data for a 124 $\mu$m two bar MoCu TES fabricated by our group. The parameters for the noise fits were measured in the usual way, with the thermal conductance parameters extracted from power law fits, and the heat capacity and $\alpha_I$ and $\beta_I$ extracted from fits to complex impedance. The factor multiplied by voltage noise $(4 k_B T R)$ to fit the measured data is shown in Fig.\ \ref{fig:r11_noise} as function of bias point. Eq.\ \ref{noise_R} once again does a good job predicting the excess noise, despite only having a limited number bias points to estimate $\partial R_d / \partial R$. The need for more densely sampled data is especially apparent at the top of noise spike between 19 \% $R_n$ and 20.5 \% $R_n$ where both the measured $\beta_I$ and excess noise peak.

Equation \ref{noise_R} needs to be carefully compared to a much larger variety of TESs. However, the two TESs discussed here are excellent examples of the types of devices where $(1+2 \beta_I)$ term fails to predict the observed noise. Although Eq.\ \ref{noise_R} does not perfectly predict the measured excess noise, the agreement is impressive given the approximations used and the uncertainty in the measurements.  In the literature, there are a few examples of TESs that show $M=0$ across most of the transition. However, these are the exception rather then the rule. They usually correspond to devices that have $\beta_I$ less than one. TESs with low $\beta_I$ are usually consistent with a dynamic resistance that is approximately linear as a function of current at a fixed temperature and well described by the two-fluid model. Then, for small values of $\beta_I$ Eq.\ \ref{noise_TF} is consistent with $(1+2 \beta_I)$. Eq.\ \ref{noise_R} could also explain measurements where dramatically different levels of excess noise are observed for the same $\beta_I$ \cite{jethava2009dependence} since the amount of noise mixed down from higher frequencies depends not just on $\beta_I$ but also on how $\beta_I$ is changing across the transition.

Unexplained excess noise has been a major impediment to designing TESs that achieve the best possible energy resolution and noise equivalent power. Using the measured dynamic resistance for a given TES and including the noise mixed down from frequencies near the Josephson frequency into the noise model for TESs are important steps to explaining the excess noise and realizing the full potential of TESs.

\begin{acknowledgments}
\end{acknowledgments}

\bibliography{m_noise_01}



\end{document}